\documentclass[10pt,twocolumn,prb,aps,showpacs,superscriptaddress]{revtex4}
\usepackage[latin9]{inputenc}
\usepackage{color}
\usepackage{units}
\usepackage{amsmath}
\usepackage{amssymb}
\usepackage{graphicx}

\makeatletter

\providecommand{\tabularnewline}{\\}

\@ifundefined{textcolor}{}
{%
 \definecolor{BLACK}{gray}{0}
 \definecolor{WHITE}{gray}{1}
 \definecolor{RED}{rgb}{1,0,0}
 \definecolor{GREEN}{rgb}{0,1,0}
 \definecolor{BLUE}{rgb}{0,0,1}
 \definecolor{CYAN}{cmyk}{1,0,0,0}
 \definecolor{MAGENTA}{cmyk}{0,1,0,0}
 \definecolor{YELLOW}{cmyk}{0,0,1,0}
}

\usepackage{graphics}
\usepackage{amsfonts}\usepackage{ulem}\usepackage{soul}\usepackage{color}

\makeatother

\begin{document}

\title{Possible spin-orbit driven spin-liquid ground state in the double
perovskite phase of Ba$_{3}$YIr$_{2}$O$_{9}$}

\author{Tusharkanti Dey }

\altaffiliation[Present address: ]{Leibniz-Institute for Solid State and Materials Research, IFW Dresden, 01171 Dresden, Germany}

\affiliation{Department of Physics, Indian Institute of Technology Bombay, Powai,
Mumbai 400076, India}

\author{A.V. Mahajan}

\email{mahajan@phy.iitb.ac.in}

\affiliation{Department of Physics, Indian Institute of Technology Bombay, Powai,
Mumbai 400076, India}

\author{R. Kumar }

\affiliation{Department of Physics, Indian Institute of Technology Bombay, Powai,
Mumbai 400076, India}

\author{B. Koteswararao}

\affiliation{Center for Condensed Matter Sciences, National Taiwan University,
Taipei 10617, Taiwan}

\author{F. C. Chou}

\affiliation{Center for Condensed Matter Sciences, National Taiwan University,
Taipei 10617, Taiwan}

\author{A. A. Omrani}

\affiliation{Laboratory for Quantum Magnetism, Ecole Polytechnique Federale de
Lausanne (EPFL), CH 1015, Switzerland}

\author{H. M. Ronnow}

\affiliation{Laboratory for Quantum Magnetism, Ecole Polytechnique Federale de
Lausanne (EPFL), CH 1015, Switzerland}
\begin{abstract}
We report the structural transformation of hexagonal Ba$_{3}$YIr$_{2}$O$_{9}$
to a cubic double perovskite form (stable in ambient conditions) under
an applied pressure of $8$\,GPa at $1273$\,K. While the ambient
pressure (AP) synthesized sample undergoes long-range magnetic ordering
at $\sim4$\,K, the high-pressure (HP) synthesized sample does not
order down to $2$\,K as evidenced from our susceptibility, heat
capacity and nuclear magnetic resonance (NMR) measurements. Further,
for the HP sample, our heat capacity data have the form $\gamma T+\beta T^{3}$
in the temperature ($T$) range of $2-10$\,K with the Sommerfeld
coefficient $\gamma=10$\,mJ/mol-Ir K$^{2}$. The $^{89}$Y NMR shift
has no $T$-dependence in the range of $4-120$\,K and its spin-lattice
relaxation rate varies linearly with $T$ in the range of $8-45$\,K
(above which it is $T$-independent). Resistance measurements of both
the samples confirm that they are semiconducting. Our data provide
evidence for the formation of a $5d$ based, gapless, quantum spin-liquid
(QSL) in the cubic (HP) phase of Ba$_{3}$YIr$_{2}$O$_{9}$. In this
picture, the $\gamma T$ term in the heat capacity and the linear
variation of $^{89}$Y $1/T_{1}$ arises from excitations out of a
spinon Fermi surface. Our findings lend credence to the theoretical
suggestion {[}G. Chen, R. Pereira, and L. Balents, Phys. Rev. B \textbf{82},
174440 (2010){]} that strong spin-orbit coupling can enhance quantum
fluctuations and lead to a QSL state in the double perovskite lattice. 
\end{abstract}


\maketitle
The $3d$ transition metal oxides have been studied extensively to
explore novel properties such as superconductivity \cite{Bednorz-ZPBCM-64-1986,Kamihara-JACS-130-2008},
colossal magnetoresistance \cite{Tokura-Colossal-2000} etc. In these
materials, the orbital angular momentum of the valence electrons is
largely quenched and a large onsite Coulomb energy often drives the
materials to a Mott insulating state. In contrast, for $5d$ group
based materials, this onsite Coulomb energy is smaller by an order
of magnitude and one could expect more metallic and less magnetic
behavior. However, some $5d$ Ir-based materials such as Na$_{4}$Ir$_{3}$O$_{8}$
\cite{Okamoto-PRL-99-2007}, Sr$_{2}$IrO$_{4}$ \cite{Kim-PRL-101-2008,Kim-Science-323-2009},
and Na$_{2}$IrO$_{3}$ \cite{Singh-PRB-82-2010} are found to be
insulators showing exotic magnetic properties. These materials are
insulating due to the presence of strong spin-orbit interactions and
are often called spin-orbit driven Mott insulators. Such materials
are relatively less explored and expected to offer a promising playground
for materials researchers.

Earlier we have investigated hexagonal Ba$_{3}$IrTi$_{2}$O$_{9}$
and suggested it to be a spin-orbit driven liquid based on a triangular
lattice \cite{Dey-PRB-2012}. Recently, Ba$_{3}$IrTi$_{2}$O$_{9}$
has been proposed as a candidate material to study Heisenberg-Kitaev
model on a triangular lattice \cite{Trebst-invited talk}. It will
be interesting to explore other iridates having a triangular lattice.
Ba$_{3}$YIr$_{2}$O$_{9}$ has a similar chemical formula like Ba$_{3}$IrTi$_{2}$O$_{9}$
and it crystallizes in the hexagonal structure (P6$_{3}$/mmc) with
Ir-Ir structural dimers arranged in an edge shared triangular fashion
\cite{Doi-JPCM-16-2004}. Since all the Ir are equivalent, they should
have a fractional oxidation state of $+4.5$ in a simple ionic picture.
Our investigation of this 5$d$-based system is motivated by the fact
that the fractional valence coupled with a geometrically frustrated
lattice might lead to a spin-liquid state or possibly a heavy fermion
state as in the 3$d$-based LiV$_{2}$O$_{4}$ \cite{Kondo-PRL-78-1997}.
This however did not turn out to be the case. Whereas we confirmed
the onset of long-range order below $4$\,K in Ba$_{3}$YIr$_{2}$O$_{9}$
(in agreement with Ref. \cite{Doi-JPCM-16-2004}), we succeeded in
suppressing the magnetic order with the application of pressure. In
fact, when Ba$_{3}$YIr$_{2}$O$_{9}$ was subjected to a pressure
of $8$\,GPa at $1273$\,K, it transformed to a cubic double perovskite
structure as evidenced from x-ray diffraction under ambient conditions.
Though the HP-synthesized sample remains insulating (based on our
resistivity measurements), it has a metal-like linear heat capacity
coefficient $\gamma=10$\,mJ/mol-Ir K$^{2}$. Further, the $^{89}$Y
NMR shift is found to be independent of temperature ($T$) below $120$\,K
and the $^{89}$Y NMR spin-lattice relaxation rate crosses over from
$T$-independent behavior at high temperature to a linear $T$-dependence
below about $45$\,K. These results point to the existence of low-energy
excitations at low-temperatures. In the absence of metallic behavior
in the resistivity and the presence of local moments, the low-$T$
data suggest the formation of an exotic ground state for the HP phase,
possibly a gapless, quantum spin-liquid (QSL) state. The ``strong
SOC'' route as opposed to the ``geometrical frustration'' route
has been suggested as a means of enhancing quantum fluctuations and
the possible formation of a QSL in, say, double perovskites \cite{Balents-PRB-2010}.
The formation of a ``spin-orbital'' liquid has been suggested earlier
in the Fe-based cubic spinel FeSc$_{2}$S$_{4}$ where the magnetic
atoms reside on the unfrustrated ``A'' sublattice \cite{Buttgen-NJP-2004,Fritsch-PRL-2004,Krimmel-PRL-2005}. 

We have prepared a polycrystalline sample of Ba$_{3}$YIr$_{2}$O$_{9}$
by solid state reaction method as detailed in Ref. \cite{Doi-JPCM-16-2004}.
Further, the AP sample was treated under $8$\,GPa pressure at $1273$\,K
for $30$\,min to get the HP sample. Due to the small size of the
HP cell, only about $150$\,mg of sample was obtained. Powder x-ray
diffraction (XRD) measurements were performed at room temperature
with Cu $K_{\alpha}$ radiation ($\lambda=1.54182\textrm{\AA}$) in
a PANalytical X'Pert PRO diffractometer. Magnetization measurements
were carried out in the temperature range $2-400$\,K and field range
$0-7$\,T using a Quantum Design SQUID VSM. Heat capacity measurements
were performed using the heat capacity attachment of a Quantum Design
Physical Properties Measurement System (PPMS) in the temperature range
$1.8-300$\,K and field range $0-9$\,T. \textcolor{black}{By breaking
a piece of hard pellet for both the AP and HP phase, we were able
to find a few single crystal-like pieces with approximate size $600\mu$m$\times$$150\mu$m$\times$$100\mu$m.
These pieces (later mentioned as single crystals) were harder than
the rest of the pellet and looked shiny under the microscope. We have
performed resistivity measurements on these single crystals using
a home built setup.} For $^{89}$Y NMR measurements, a Tecmag pulse
spectrometer and a fixed magnetic field $93.954$\,kOe obtained inside
a room-temperature-bore Varian superconducting magnet were used. Variable
temperature was obtained with the help of an Oxford cryostat and accessories.
The $^{89}$Y nucleus has spin $I=1/2$ ($100\%$ natural abundance)
and gyromagnetic ratio $\gamma/2\pi$ = $2.08583$\,MHz/T.

\begin{figure}
\centering{}\includegraphics[scale=0.35]{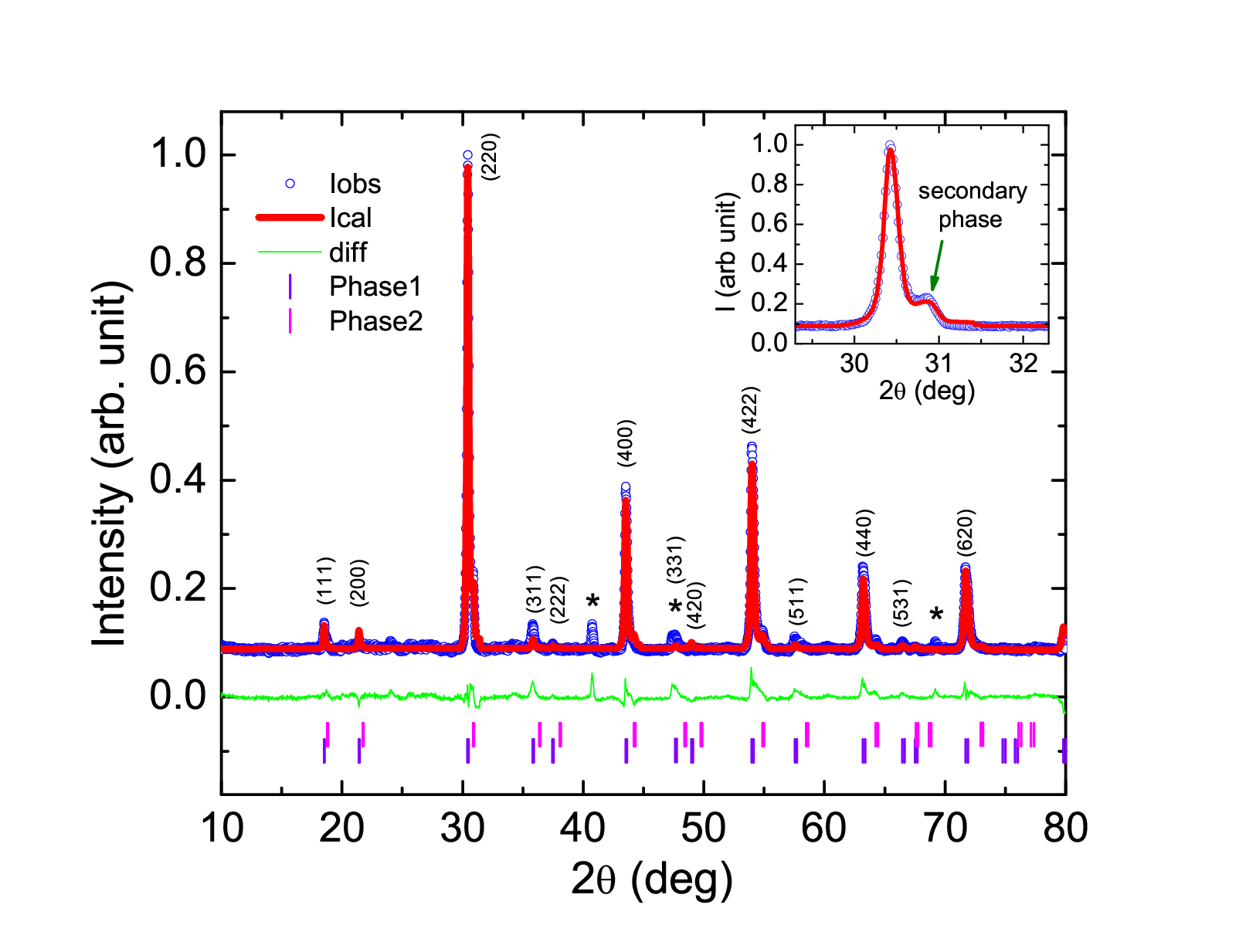}\caption{\label{fig:XRD}The x-ray diffraction pattern of the HP phase of Ba$_{3}$YIr$_{2}$O$_{9}$
sample is shown along with a two phase refinement with space group
Fm-$3$m. The (hkl) positions are also indicated. A $5\%$ impurity
phase of unreacted Ir is found. The peaks corresponding to Ir are
marked with ({*}). Inset: The main peak of the primary phase is shown.}
\end{figure}

\begin{table}
\centering{}\caption{\label{tab:XYZ positions}Atomic parameters obtained from refinment
of x-ray powder diffraction pattern at room temperature for the primary
phase of the HP sample. }
\begin{tabular}{|c|c|c|c|c|c|}
\hline 
Atoms & Site & x & y & z & Occupancy\tabularnewline
\hline 
Ba(1) & 8c & 0.25 & 0.25 & 0.25 & 1\tabularnewline
\hline 
Y(1) & 4a & 0 & 0 & 0 & 0.67\tabularnewline
\hline 
Ir(1) & 4a & 0 & 0 & 0 & 0.33\tabularnewline
\hline 
Ir(2) & 4b & 0.5 & 0.5 & 0.5 & 1\tabularnewline
\hline 
O(1) & 24e & 0.2615(4) & 0 & 0 & 1\tabularnewline
\hline 
\end{tabular}
\end{table}

\begin{figure}
\centering{}\includegraphics[scale=0.65]{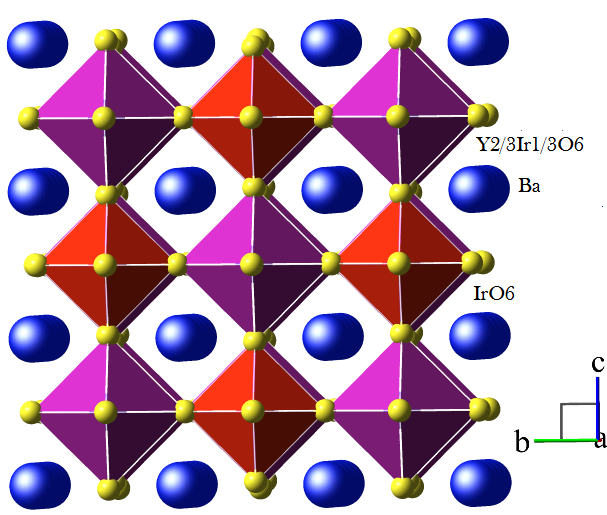}\caption{\label{fig:Structure}The crystal structure of the high pressure synthesized
cubic phase is shown. The blue and yellow atoms represent Ba and O,
respectively. The red octahedra are IrO$_{6}$ octahedra while the
violet octahedra contain Ir or Y at their centers.}
\end{figure}

Our XRD analysis confirmed that the AP sample was formed in a hexagonal
structure (P6$_{3}$/mmc) as reported earlier \cite{Doi-JPCM-16-2004}
but we found that after high-pressure treatment the structure changed
to a cubic double perovskite (Ba$_{2}$MM'O$_{6}$) phase with space
group Fm-$3$m. It is relevant to mention that Ba$_{3}$NiSb$_{2}$O$_{9}$
was also reported to transform from a hexagonal (P6$_{3}$/mmc) to
cubic (Fm-$3$m) phase when treated under $9$\,GPa pressure at $1273$\,K
for $30$\,min \cite{Cheng-PRL-107-2011}. In the XRD pattern (Fig.
\ref{fig:XRD}) of the HP phase, a few extra peaks (marked as {*})
are present which are identified as due to unreacted Ir and account
for about $5\%$ of the contribution. Also, all the major peaks have
a shoulder on the right side as shown in the inset of Fig. \ref{fig:XRD}.
These shoulders possibly arise from a double perovskite phase with
a slightly different cell parameter (this could happen due to a miscibility
gap in the phase diagram). Accordingly, a two-phase Rietveld refinement
(Fig. \ref{fig:XRD}) performed with our XRD data yields lattice parameter
$a=8.3248\text{\AA}$ for the primary phase and $a'=8.202$$\text{\AA}$
for the secondary phase (about $10\%$ content). A similar double
perovskite Ba$_{2}$YIrO$_{6}$ (space group Fm-$3$m) has lattice
parameter $a=8.35032\text{\AA}$\cite{Fu-JAlloyComp-394-2005}. Atomic
cell parameters for the primary phase resulting from our refinement
are listed in Table \ref{tab:XYZ positions}. Attempts to explain
the XRD data using a cubic perovskite structure (chemical formula
Ba(Y$_{1/3}$Ir$_{2/3}$)O$_{3}$) with approximately half the lattice
parameter of the double perovskite (chemical formula Ba$_{2}$Ir(Y$_{2/3}$Ir$_{1/3}$)O$_{6}$)
did not lead to a good refinement. On the other hand, we find that
an ordered arrangement of IrO$_{6}$ and Y$_{2/3}$Ir$_{1/3}$O$_{6}$
octahedra form the cubic double perovskite Ba$_{2}$Ir(Y$_{2/3}$Ir$_{1/3}$)O$_{6}$
structure of the HP phase (shown in Fig. \ref{fig:Structure}). We
note that every Y will have $6$ Ir as its nearest neighbors while
as next nearest neighbors it will have $33\%$ Ir and $67\%$ Y.

\begin{figure*}
\raggedright{}\centering{}\includegraphics[scale=0.66]{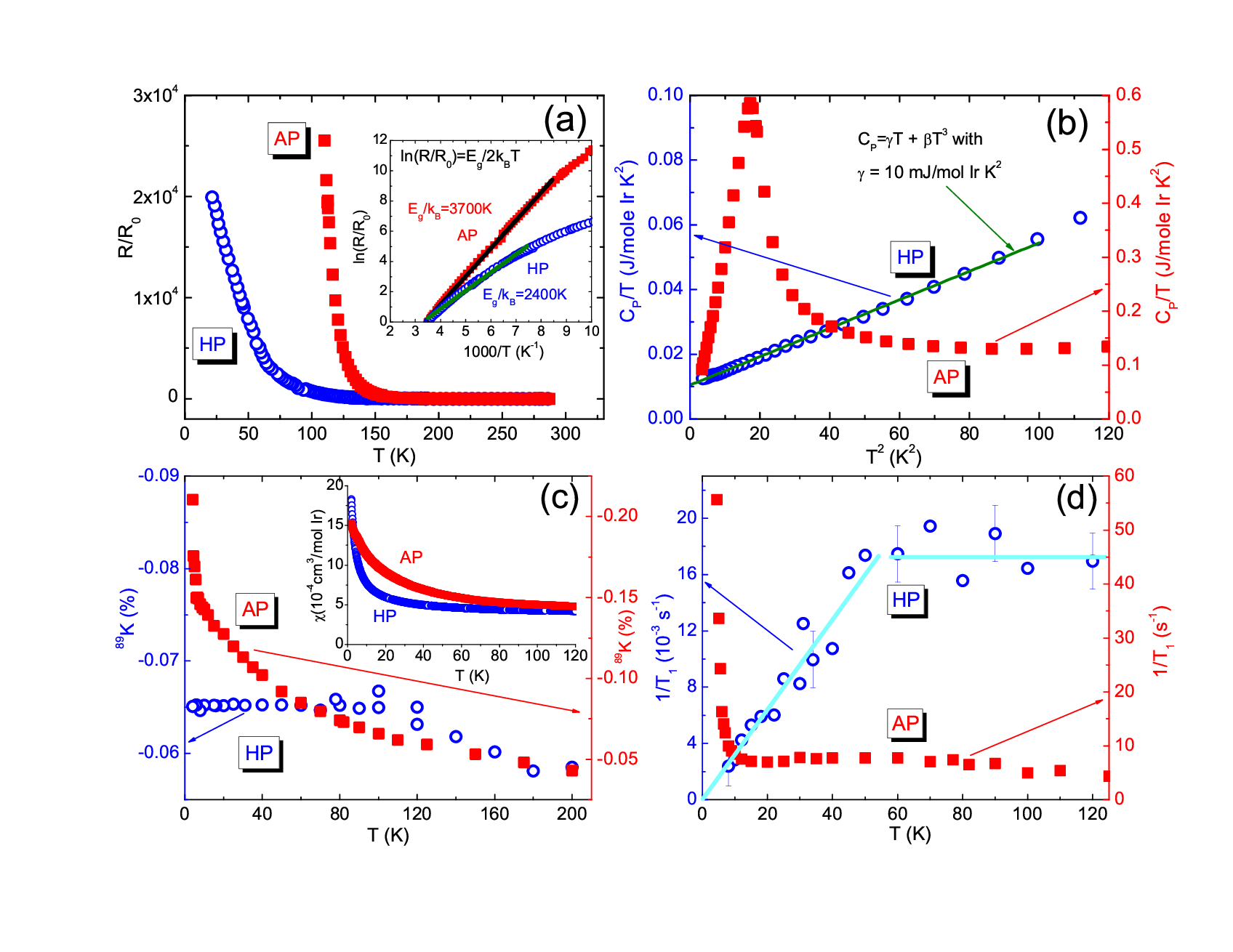}\caption{\label{fig:CombinedPanel}The results of different measurements on
the HP sample (blue open circles) are shown in comparison with its
AP counterpart (red solid squares). In case of (b), (c) and (d), AP
sample corresponds to the right axis and HP sample to the left axis
while for (a) both correspond to the left axis. (a) The normalized
resistivity is shown as a function of temperature. Inset: ln(R/R$_{0}$)
is plotted with $1000/T$. The solid lines are fit with the formula
mentioned in the figure. (b) The $C_{P}/T$ is shown as a function
of $T^{2}$. The green solid line is a fit of the HP data with the
formula stated in the figure. (c) Temperature variation of $^{89}$Y
NMR shift for the HP and AP samples are shown. Their bulk susceptibilities
are shown in the inset. (d) The $^{89}$Y spin-lattice relaxation
rate ($1/T_{1}$) is shown as a function of temperature. The light
blue solid lines are guides to eye.}
\end{figure*}

\begin{figure}
\centering{}\includegraphics[scale=0.35]{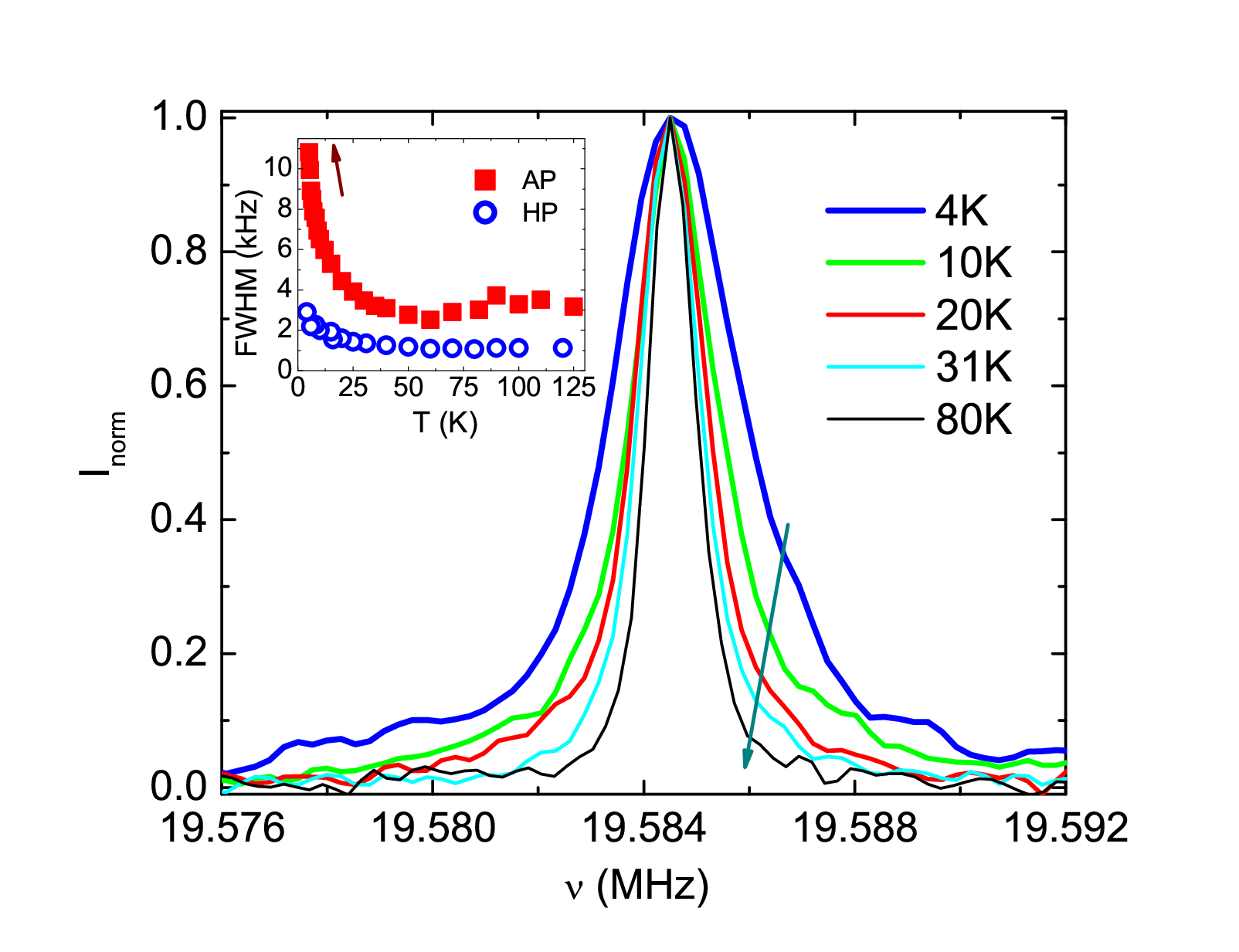}\caption{\label{fig:Spectra} $^{89}$Y NMR spectra at different temperatures
for the HP sample are shown. The arrow indicates measurement temperature
in increasing order. Inset: FWHM of the AP sample as well as the HP
sample as functions of temperature. The arrow indicates that FWHM
for the AP sample goes much higher at low temperature ( $40$\,kHz
at $4$\,K).}
\end{figure}

Figure \ref{fig:CombinedPanel} provides an overview of our data pertaining
to some basic measurements on HP and AP phases of Ba$_{3}$YIr$_{2}$O$_{9}$.
Resistivity data (shown in Fig. \ref{fig:CombinedPanel}(a)) on single
crystals show insulating behavior for both AP and HP samples. The
AP sample shows an activated behavior with a gap of about $0.3$\,eV.
For the HP sample, as well, the resistivity increases with decreasing
temperature though with a weaker rise. In the heat capacity, on the
other hand, a clear anomaly at about $4$\,K is seen for the AP sample
(Fig. \ref{fig:CombinedPanel}(b)) which suggests long-range order.
The anomaly goes away in the HP sample indicating that the ordering
has been suppressed by the application of pressure. It is further
interesting to note that for the HP sample below about $10$\,K,
the heat capacity $C_{P}(T)$ is metal-like with a linear $C_{P}/T$
vs $T^{2}$ (with slope $\beta=0.44$\,mJ/mol-Ir K$^{4}$) and a
non-zero $y$-axis intercept giving a Sommerfeld coefficient $\gamma=10$\,mJ/mole-Ir
K$^{2}$. The inset of Fig. \ref{fig:CombinedPanel} (c) shows the
magnetic susceptibility $\chi(T)$ of the samples in a field of $5$\,kOe.
For the AP sample, magnetic susceptibility shows a kink at $T\sim4$\,K
indicative of magnetic ordering as reported earlier \cite{Doi-JPCM-16-2004}.
Susceptibility data of the AP sample could be fitted to the Curie
law (with $\theta\sim0$) in the temperature range $30-300$\,K which
yields a $T$-independent susceptibility $\chi_{0}=3.8\times10^{-4}$\,cm$^{3}$/mol
Ir and a Curie constant $C=0.0125$\,cm$^{3}$K/mol Ir. The small
Curie term (nearly thirty times smaller than that for $S=1/2$) is
not unusual in iridates. In fact many other iridates have been reported
with small Curie terms \cite{Dey-PRB-2012,Cao-PRB-57-1998-Sr2IrO4,Baker-PRB-87-2013}.
Below about $30$\,K, the susceptibility deviates from Curie behavior
presumably due to the building up of magnetic correlations. Consistent
with the susceptibility behavior in the AP sample, the $^{89}$Y NOR
shift ($K$) increases with a decrease in temperature (Fig. \ref{fig:CombinedPanel}
(c)). One should note that whereas in the AP phase the $\theta_{\mathrm{CW}}$
is nearly zero from $\chi(T)$ data, a CW fit of $K(T)$ gives $\theta_{\mathrm{CW}}\approx-30$K
\cite{Dey-Ba3Y-ScIr2O9}. For the HP sample, on the other hand, any
susceptibility anomaly indicative of a magnetic ordering was not seen
down to $2$\,K nor was there any ZFC-FC splitting observed. The
$\chi(T)$ for the HP sample could be fitted to the CW law in almost
the whole temperature range resulting in the parameters $\chi_{0}=3.9\times10^{-4}$\,cm$^{3}$/mol
Ir, $C=0.0045$\,cm$^{3}$K/mol Ir and $\theta=-1.6$\,K. For the
HP sample, the $C$ value is smaller by a further factor of $3$ compared
to its AP counterpart while the $\chi_{0}$ is nearly unchanged. We
have measured $^{89}$Y NMR spectra for the HP sample at different
temperatures as shown in Fig. \ref{fig:Spectra}. The spectra for
the HP sample is very narrow ($\sim2$\,kHz) and becomes broader
with decreasing temperature. The full width at half maxima (FWHM)
for the HP sample is shown as a function of temperature in the inset
of Fig. \ref{fig:Structure} and compared with the FWHM for the AP
sample. The temperature variation of FWHM for the HP sample is much
weaker in comparison to the AP sample. However the increase in FWHM
at low temperature probably suggests presence of small local moments
in the HP sample. As shown in Fig. \ref{fig:CombinedPanel}(c), the
$^{89}$Y NMR shift (obtained from the peak position of individual
spectra) for the HP sample increases with a decrease in temperature
and becomes temperature independent below about $100$\,K. The magnitude
of the change in $K$ with temperature is five times smaller than
for the AP sample. These points indicate that the magnetic moments
in the HP sample are very weak. Once again, the small $\theta_{\mathrm{CW}}$
from $\chi(T)$ analysis need not suggest weak correlations but might
simply suggest that the intrinsic moments are very small and that
the bulk susceptibility data are dominated by extrinsic effects. The
saturation of $K$ below $100$\,K suggests a quenching of the local
moments and the small low-$T$ Curie term seen in the bulk susceptibility
is likely extrinsic.

\begin{figure}
\centering{}\includegraphics[scale=0.35]{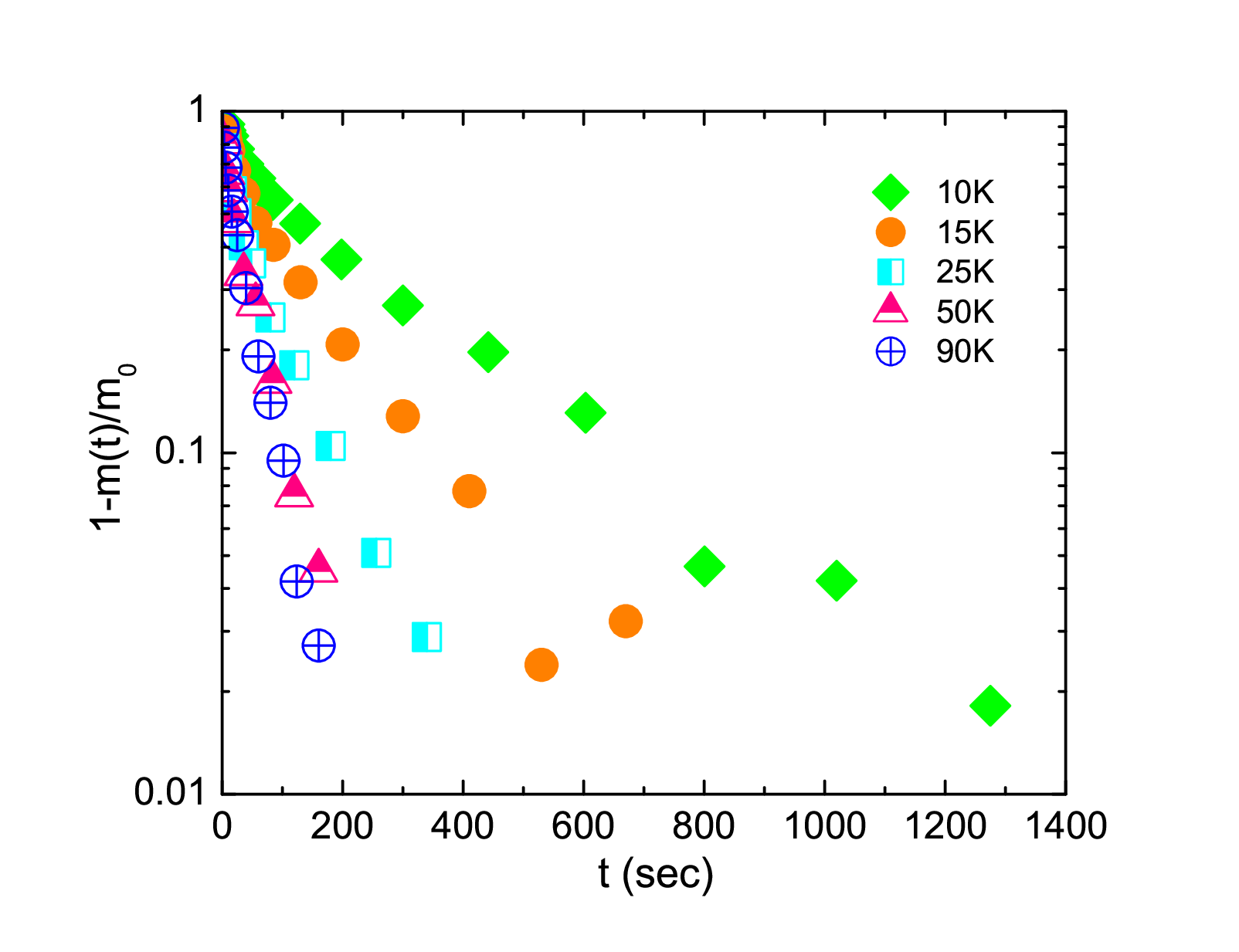}\caption{\label{fig:Recoveries} Shown here are some representative data for
the HP sample for the recovery of the $^{89}$Y longitudinal nuclear
magnetization after a saturating pulse.}
\end{figure}

We next probe the low-energy excitations of both the AP and the HP
sample using $^{89}$Y NMR spin-lattice relaxation rate measurements.
We have measured the spin-lattice relaxation rate ($1/T_{1}$) using
a standard saturation recovery method following a $\nicefrac{\pi}{2}-t-(\nicefrac{\pi}{2}-\pi)$
pulse sequence in the temperature range $4-300$\,K for the AP sample
and $8-120$\,K for the HP sample. For the AP sample, nuclear magnetization
recoveries are single exponential as expected for an $I=1/2$ nucleus
\cite{Dey-Ba3Y-ScIr2O9}. The relaxation time ($T_{1}$) for the AP
sample varies from about $10$\,ms to about $400$\,ms as a function
of temperature. For the HP sample, the nuclear magnetisation recovery
after a saturating pulse has an initial short component followed by
a longer component. A few representative recovery data are shown in
the Fig. \ref{fig:Recoveries}. We have fitted them with the formula
\begin{equation}
1-m(t)/m_{0}=A[Bexp(-t/T_{L})+(1-B)exp(-t/T_{S})]\label{eq:T1 recovery}
\end{equation}
where $T_{L}$ and $T_{S}$ are the longer and shorter components
of $T_{1}$, respectively, and $B$ is $\sim0.7$. The longer component
$T_{L}$ varies from about $40$\,s at $100$\,K to about $400$\,s
at $8$\,K. Below $8$\,K, $T_{L}$ becomes even longer though we
did not take detailed data. The major differences of relaxation rate
($1/T_{1}$) between the AP and HP sample are, (i) $1/T_{1}$ is almost
three orders of magnitude smaller for the HP sample compared to the
AP sample and (ii) $1/T_{1}$ increases with decreasing temperature
for the AP sample but decreases with decreasing temperature ($T<45$\,K)
for the HP sample. A clear signature of ordering is seen in the divergence
of the $^{89}$Y NMR $1/T_{1}$ for the AP sample (Fig. \ref{fig:CombinedPanel}
(d)). In contrast, for the HP sample, $1/T_{1}$ is $T$-independent
at high-$T$ as might be expected in a paramagnetic insulator but
crosses over to a linear variation below about $45$\,K (the rate
corresponding to the longer $T_{1}$ component ($1/T_{L}$) is plotted
here).

To summarize, whereas the AP sample shows unambiguous evidence of
being a magnetic insulator, the data on the HP sample gives apparently
conflicting signals. Whereas the low-$T$ NMR shift (and hence the
intrinsic susceptibility), heat capacity, and $1/T_{1}$ are all metal-like,
the resistivity (measured on a single crystal) is insulating! We suggest
that these results imply the exotic possibility of the absence of
charge excitations while spin-excitations are present. The linear
contribution to the low-$T$ heat capacity as also to the $^{89}$Y
$1/T_{1}$ then come from excitations from a spinon Fermi surface
as might happen in a gapless QSL. In this context it is relevant to
point out NMR results on other spin liquid systems. For the \textbf{$5d$-}based
spin liquid system Na$_{4}$Ir$_{3}$O$_{8}$, $^{23}$Na NMR by Takagi
\textit{et al.} \cite{Takagi-PPT} finds a negligible temperature
dependence of shift below about $100$\,K reflecting the independence
from temperature of the intrinsic susceptibility. In this case the
$^{23}$Na $1/T_{1}$ was constant above $200$\,K below which it
followed a power law with power $1$ in the range $2-10$\,K and
with power $2.5$ in the temperature range $100-200$\,K\textit{.}
They attributed this low-temperature linear behavior to low lying
spin excitations. Itou \textit{et al.} \cite{Itou-PRB-84-2011} also
found almost no temperature dependence of shift in their $^{13}$C
NMR data down to $18.7$\,mK on the organic system EtMe$_{3}$Sb{[}Pd(dmit)$_{2}${]}$_{2}$
with a spin liquid ground state while $1/T_{1}$ was found to be constant
in some range above $1$\,K and varied as $T^{2}$ below $1$\,K.
In the organic spin liquid system $\kappa$-(BEDT-TTF)$_{2}$Cu$_{2}$(CN)$_{3}$,
Shimizu \textit{et al.} \cite{Shimizu-PRL-91-2003} found no change
in $^{1}$H NMR spectral position from $36.1$\,K down to $32$\,mK
while $1/T_{1}$ followed a power law at low-temperature, from which
the authors concluded that the ground state has gapless low lying
spin excitations. Our NMR data are qualitatively similar to the cases
listed above, providing further justification for the spin liquid
state at low $T$. It is worth mentioning that another ordered double
perovskite Ba$_{2}$YMoO$_{6}$ was reported to have an exotic valence
bond glass state \cite{deVries-PRL-104-2010-Ba2YMoO6} or a collective
(gapped) spin-singlet state \cite{Aharen-PRB-81-2010-Ba2YMoO6NMR}
whereas Ref. \cite{Balents-PRB-2010} suggested a quadrupolar ordered
state.

In conclusion, we have presented evidence based on bulk probes (resistivity,
susceptibility, heat capacity) and local probes ($^{89}$Y NMR shift
and $1/T_{1}$) for a possible gapless QSL state in the HP (cubic
double perovskite) phase of Ba$_{3}$YIr$_{2}$O$_{9}$. The occurrence
of a spin-liquid state even for a geometrically non-frustrated lattice
suggests the presence of anisotropic exchange interactions strongly
tied to the large spin-orbit coupling in this $5d$-based system.
The AP phase on the other hand was found to order magnetically around
$4$\,K. A strong SOC likely plays a crucial role in the insulating
behavior of both AP and HP samples. 

We thank the Department of Science and Technology, Govt. of India
for financial support. This work was partially supported by Swiss
National Science Foundation and the Indo Swiss Joint Research Programme.
FCC acknowledges the support from National Science Council of Taiwan.
We thank B. H. Chen for his help in preparation of the HP phase sample.
We also thank S. K. Panda, I. Dasgupta, Kedar Damle and P. P. Singh
for useful discussions.

\end{document}